\documentclass[conference, a4paper]{IEEEtran}

\IEEEoverridecommandlockouts
\usepackage{booktabs}
\usepackage{graphicx}
\usepackage{multirow}
\usepackage{amsmath,amssymb,latexsym}
\usepackage{rotating}
\usepackage{lettrine}
\usepackage{textcomp}
\usepackage{units}
\usepackage[table]{xcolor}
\usepackage{alltt}
\usepackage{hyperref}
\usepackage{tikz}
\usepackage{lipsum}%
\usepackage[pscoord]{eso-pic}
\newcommand{\placetextbox}[3]{%
  \setbox0=\hbox{#3}
  \AddToShipoutPictureFG*{
    \put(\LenToUnit{#1\paperwidth},\LenToUnit{#2\paperheight}){\vtop{{\null}\makebox[0pt][c]{#3}}}%
  }%
}%

\usepackage{xcolor}

\title{Executable QR codes with Machine Learning for Industrial Applications
\thanks{This work has been partially funded by CNR under the project ``Executable QR codes (EQR) - GORU IEIIT'' (DIT.AD001.212).}
}

\author{
    \IEEEauthorblockN{
    Stefano Scanzio\IEEEauthorrefmark{1},
    Francesco Velluto\IEEEauthorrefmark{2},
    Matteo Rosani\IEEEauthorrefmark{1}\IEEEauthorrefmark{2},
    Lukasz Wisniewski\IEEEauthorrefmark{3},
    Gianluca Cena\IEEEauthorrefmark{1}
    }
    
    \IEEEauthorblockA{\IEEEauthorrefmark{1}National Research Council of Italy (CNR--IEIIT), Italy. \IEEEauthorrefmark{2}Politecnico di Torino, Italy.}    
    \IEEEauthorblockA{\IEEEauthorrefmark{3}Institute Industrial IT - inIT, Technische Hochschule OWL, Germany.}
    Emails: stefano.scanzio@cnr.it, vellfr00@gmail.com, matteo.rosani@polito.it,\\lukasz.wisniewski@th-owl.de, gianluca.cena@cnr.it
    }

\begin{document}
\placetextbox{0.5}{1}{This is the author's version of an article that has been published.}
\placetextbox{0.5}{0.985}{Changes were made to this version by the publisher prior to publication.}
\placetextbox{0.5}{0.97}{The final version of record is available at \href{https://doi.org/10.1109/ETFA61755.2024.10710739}{https://doi.org/10.1109/ETFA61755.2024.10710739}}%
\placetextbox{0.5}{0.05}{Copyright (c) 2024 IEEE. Personal use is permitted.}
\placetextbox{0.5}{0.035}{For any other purposes, permission must be obtained from the IEEE by emailing pubs-permissions@ieee.org.}%

\maketitle
\thispagestyle{empty}
\pagestyle{empty}

\begin{abstract}
Executable QR codes, also known as eQR codes or just sQRy, are a special kind of QR codes that embed programs conceived to run on mobile devices like smartphones. Since the program is directly encoded in binary form within the QR code, it can be executed even when the reading device is not provided with Internet access.
The applications of this technology are manifold, and range from smart user guides to advisory systems. 
The first programming language made available for eQR is QRtree, which enables the implementation of decision trees aimed, for example, at guiding the user in operating/maintaining a complex machinery or for reaching a specific location.

In this work, an additional language is proposed, we term QRind, which was specifically devised for Industry. It permits to integrate distinct computational blocks into the QR code, e.g., machine learning models to enable predictive maintenance and algorithms to ease machinery usage. 
QRind permits the Industry 4.0/5.0 paradigms to be implemented, in part, also in those cases where Internet is unavailable.
\end{abstract}


\section{Introduction}
\label{sec:introduction}
Interaction between humans and objects in today's real world, as postulated by the Internet of Things (IoT) paradigm, is increasingly based on the use of QR codes, and the industrial context is no stranger to this trend. 
Since their creation in 1994, QR codes have been commonly employed to 
produce labels that
quickly provide various types of information,
e.g., to track the products' lifecycle.
In the following three decades, QR code technology has proven to be extremely pervasive and has been successfully applied to a large number of different fields, to cite a few: localization \cite{10500827}, payment \cite{42}, recycling \cite{46}, security \cite{10}, teaching \cite{33}, traceability \cite{43}, etc.

Nowadays, QR codes are often used to store a uniform resource identifier (URI), or more specifically a uniform resource locator (URL) that points to a remote resource (e.g., a web server) on which either useful information is found or some web-based application is run.
Executable QR codes (eQR codes), also known as sQRy, have been introduced in 2022 with the aim to overcome one of the major limitations of conventional QR codes, 
that is, their inability to support user interaction
in the absence of a working Internet connection. 

In eQR codes, the application is completely embedded inside the QR code (encoded as a binary code),
making execution possible even when network access is not available. 
This situation is not uncommon in real life (e.g., when the user is located inside a large building, in the mountains, in the middle of the sea), but can be encountered also in industrial scenarios. 
Consider, for instance, petrochemical plants, mines, but also shop-floors and warehouses where, for whatever reason, maintenance personnel lack authorization to connect to both the corporate intranet and the Internet.

In this work a new programming language (or dialect, in the eQR code terminology) is proposed, we name QRind. 
Similarly to QRtree, which is currently the sole existing dialect implemented in eQR codes, QRind permits the implementation of decision trees, which are useful for predictive/reactive maintenance and troubleshooting, but can be also employed to assist users in operating complex machinery.
QRind is a completely different dialect from its predecessor: 
not only it permits the declaration of variables, but enables integration of machine learning algorithms, in such a way to improve equipment maintainability as foreseen by Industry 4.0/5.0. 
At the same time, QRind retains high compactness for the generated code, which is probably the main requirement for any eQR code dialect, as QR code capacity is very limited.

In the following, Section~\ref{sec:eQR} introduces QR codes and the eQR code technology, 
Section~\ref{sec:QRind} describes the basics of the QRind dialect, 
which is further explained through an example in Section~\ref{sec:example}, 
just before the conclusive remarks of Section~\ref{sec:conc}.

\section{QR codes vs. eQR codes (sQRy)}
\label{sec:eQR}
QR code is a two-dimensional barcode technology invented by Denso Wave Automotive company 
three decades ago to track (i.e., identify) vehicles during manufacturing. 
The standard has been updated over the years to increase capacity and recognition speed, to the point where they are now far superior to one-dimensional barcodes. 
The most recent specification dates back to 2015 \cite{2015-QRcode} and defines several versions of QR codes (from $1$ to $40$), which can contain up to $2953$ bytes.

A QR code typically consists of a grid of black squares on a white background, along with some graphical control structures to aid detection and alignment of the optical reader during the decoding process.
Every QR code is generated starting from the information that we want to store in it (more information requires larger physical space).
The outcome is an image that can be printed on a physical support, like paper, and made available to users
easily and inexpensively.

Data contained in a QR code can be of four types: numerical, alphabetical, kanji/kana, or binary. 
Each one of these types has its own encoding, which results in different physical space requirements.
Error correction codes are additionally exploited to improve reliability, so that even partially damaged QR codes can be read.
Error correction takes some space, so a trade-off between reliability and net capacity is sought. 
This can be selected via four possible levels of error correction (Low, Medium, Quartile, and High), which offer increasing resiliency to damage allowing up to $30\%$ of data to be recovered.

Basically, 
an eQR code (sQRy) is a QR code that contains a program written in the \textit{QRscript} programming language \cite{scanzio2024qrscript}. 
The idea of including a program into a QR code was preliminarily proposed in \cite{9921530} and later expanded in \cite{10492995}.
eQR technology permits intelligence to be embedded inside a QR code and its operation does not require Internet access. 
The eQR code is a self-contained program that can be operated without any external resources, except for a virtual machine installed on the end-user's device, which provides a dedicated software environment that executes the binary code retrieved from the QR code \cite{scanzio2024qrtree}. 
This is a key enabler in all those contexts where the Internet connection is absent or just unstable.

To decrease the space occupied by the encoded program,
the QRscript programming language supports a plurality of \textit{dialects}.
In fact, every dialect is very specific, which implies a narrower application field and a smaller instruction set, 
which translates to a more compact encoding for instructions.
For example, the QRtree dialect \cite{10492995,scanzio2024qrtree} supports the representation of decision trees and only includes $7$ instructions.

\begin{figure}[b]
	\begin{center}
    \vspace{-0.3cm}  
	\includegraphics[width=0.84\columnwidth]{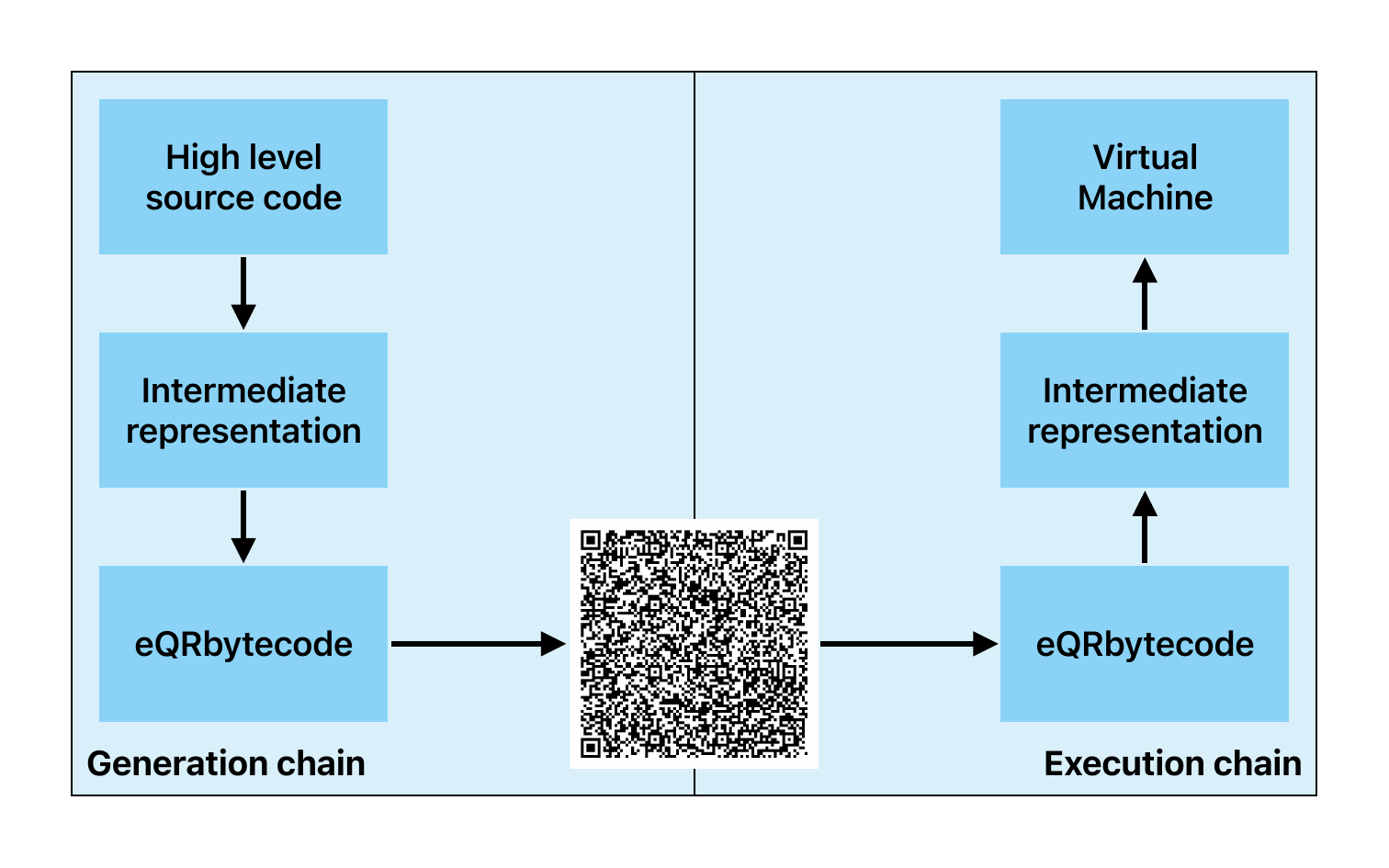}
	\end{center}

    \caption{Generation chain and execution chain for eQR codes.}
	\label{fig:generationExecutionChain}
\end{figure}

As shown in Fig.~\ref{fig:generationExecutionChain},
the life cycle of an eQR code is split in two parts, namely, the \textit{generation} and the \textit{execution} chains.

The generation chain is carried out by the eQR developer (expert user), typically on a workstation or a PC. 
First, the source code (a high-level programming language, even a visual one) is translated into an intermediate representation. 
More than one high-level language can be translated to the same intermediate representation. 
Second, the intermediate representation is converted to a binary format termed \textit{eQRbytecode}. 
Finally, the stream of bytes is embedded into a
conventional
QR code. 
Since suitable libraries are widely available for this purpose, the last step is trivial.

The execution chain is instead performed by the end user
on her/his
device (e.g., a smartphone). 
A specialized app has to first recognize and extract the 
content of the 
QR code, obtaining the related eQRbytecode. 
Again, there are plenty of libraries aimed to this purpose. 
The eQRbytecode is then converted back into the intermediate representation.
Finally, the program is executed by a virtual machine 
running on the user's device
that supports user interaction through the specific input/output features offered by the device itself (typically, a touchscreen).

\section{QRind Dialect}
\label{sec:QRind}

\begin{table*}[th]
  \caption{Main instructions defined for the intermediate representation of the QRind dialect \\ (\texttt{<reg>} represents a register, \texttt{<lit>} represents a literal). 
  }
\vspace{-0.3cm}
\label{tab:instructions}
  \footnotesize
  \begin{center}
    \tabcolsep=0.15cm	
    \renewcommand{\arraystretch}{1.1}
    \begin{tabular}{l|l|l}
    Instruction & Description & Syntax \\    
    \hline
    \texttt{SET} & Set a register & \texttt{SET <reg> TO <val>}  \\
    \texttt{INPUT} & Ask user for an input & \texttt{INPUT <type> INTO <reg>} \\
    \texttt{PRINT} & Print a register & \texttt{PRINT (<reg>|<lit>)} \\
    \texttt{TREECONDITION} & Conditional jump & \texttt{TREECONDITION (<reg>|<lit>) <operator> (<reg>|<lit>) <label>} \\
    \texttt{TREEJUMP} & Unconditional jump & \texttt{TREEJUMP <label>} \\
    \hline
    \texttt{MLINPUT} & Map a register to the ML input & \texttt{MLINPUT <ML\_type> <int> FROM <register>}  \\
    \texttt{NNLAYER} & Specify next layer & \texttt{NNLAYER <int> <type>
<encoding> <weights\_list>} \\
    \texttt{MLOUTPUT} & Map the ML output into a register & \texttt{MLOUTPUT INTO <register>} \\
    \hline
    \end{tabular}
    \vspace{-0.4cm}
        
  \end{center}
\end{table*}
This section describes the intermediate code for the QRind dialect. 
As for QRtree, the definition of a high-level programming language is irrelevant from a research viewpoint. 
Such languages only facilitate the user in writing an intermediate (optimized) language, but they are not representative of the properties of the language.

The QRind dialect shares with QRtree the ability to store a decision tree, 
which is a fundamental model to perform \textit{diagnosis} and \textit{predictive maintenance}, 
e.g., to implement an \textit{assistance system} for failure management or smart instruction manuals for industrial machinery.
Unlike QRtree, QRind supports the declaration of variables. 
This increases the size of the intermediate code slightly, but greatly enhances expressiveness and, thus, the set of concrete problems this technology can tackle.
Such feature is indispensable in a language that allows the definition of ML models, because it permits to feed the ML algorithm with a set of input features to perform classification and regression. 
Usually, input values for the ML model cannot be requested from the user immediately before the execution of the ML algorithm, 
since they may have already been acquired elsewhere in the program or obtained as the outputs of other ML or non-ML algorithms.
 
Variables are stored in registers. 
The number of registers that can be defined is potentially unlimited, because the computational resources of the device on which the algorithm is executed (a modern smartphone) exceed the amount of information that can actually fit into a QR code.
The index used to address a register exploits the \textit{exponential encoding}, which permits to encode any integer number using a variable number of bits and is described in detail in \cite{scanzio2024qrtree}.
Registers used more often are encoded on a smaller number of bits. 
This permits compact representations without imposing any limitations on the number of registers. 

In particular, the following types of variables were defined (which can be useful for the application context): \textit{boolean}, 
\textit{int} ($8$ and $16$ bit representations), 
\textit{float} ($16$ and $32$ bit representations), 
\textit{string} (ASCII-7 and UTF-8 representations), 
and \textit{arrays}. 
For the latter type, both homogeneous arrays (which can be encoded with fewer bits) and heterogeneous arrays (which provide higher versatility from an application perspective) are available. 
At the current stage of the QRind design, we plan the possibility of having other data types like dates (helpful, e.g., to manage scheduled maintenance of aging machinery) and very large numbers. 

Table~\ref{tab:instructions} reports the most common instructions of QRind. 
The first five instructions are aimed at managing the decision trees, while the latter three are intended for ML capability. 
In particular, excluding the first three instructions for assignment, input, and output, 
the instructions \texttt{TREECONDITION} and \texttt{TREEJUMP} are for conditional and unconditional jumps, respectively.  
The addition of instructions to deal with mathematical and boolean operations could be very useful from an application point of view. 
Their description has not been reported here, however, since their implementation is simple. 
Typically, it results in the execution of a sequence of operations in reverse Polish notation (in which case computation can be easily done using a stack).

More interesting and specific are the last three instructions, related to the implementation of ML algorithms. 
The \texttt{MLINPUT} instruction feeds a register (array) to 
the ML model and is used to transfer the related input features. 
The instruction identifies with \texttt{<ML\_type>} the type of the ML algorithm. 
In the example we analyzed a
multi-layer perceptron (MLP), which is one of the most common and widely used ML algorithms. 
MLP is identified by type \texttt{000}, while \texttt{<int>} represents the number of inputs to the model.

The instruction \texttt{NNLAYER} is instead specific to ANNs of type MLP. 
They are used to construct the ANN layer-by-layer, from the input to the output. 
Parameter \texttt{<int>} represents the number of neurons of the next layer, 
\texttt{<type>} is the type of neurons (the same for the entire layer, e.g., linear, sigmoid, tanh, ReLU, Leaky ReLU, softmax), 
\texttt{<encoding>} expresses how weights are encoded (float16 for more compact models or float32 for precise models), 
and finally \texttt{<weights\_list>}, which represents the list of weights and biases of the model. 
The \texttt{<weights\_list>} is encoded starting from the leftmost neuron of the upper layer by listing the weights that connect this neuron with the neurons of the lower layer, in left-to-right order, and then the corresponding bias.

The formula implementing the run forward process of one layer of the MLP is:
\begin{equation}
out_i = f \Big( \sum_{j=1}^{n_l} w_{ij}\cdot x_j + b_i \Big), \quad i \in \{1, 2, \ldots, n_{l+1}\}, \label{eq:ann}
\end{equation}
where $n_l$ is the number of neurons in the lower layer $l$, 
$n_{l+1}$ is the number of neurons in the next layer $l+1$, 
$w_{ij}$ is the weight connecting neuron $j$ in layer $l$ with neuron $i$ in layer $l+1$, 
$x_j$ is the input that is mapped to a specific register,
$b_i$ is the bias of neuron $i$ in layer $l+1$, 
$f(\bullet)$ is the activation function (linear, sigmoid, tanh, etc.), 
and $out_i$ is the output of neuron $i$ of layer $l+1$.
For example, if two adjacent layers are characterized by $n_l=2$ and $n_{l+1}=3$, the weights and biases are encoded by QRind in the following order: $w_{11},w_{12},b_{1},w_{21},w_{22},b_{2},w_{31},w_{32},b_{3}$.

Finally, the \texttt{MLOUTPUT} function maps the output of the ML model to a register, which can be used to alter the way the decision tree can be 
subsequently
navigated by the user.

Once the intermediate language instructions are defined, a set of conversion rules must be specified for translation from intermediate language to binary code, as customarily done in compilers.
While not particularly complex (an example can be found in the QRtree specification document \cite{scanzio2024qrtree}), such rules should aim to produce a code that is as compact as possible.

\section{Example}
\label{sec:example}

\begin{figure}[t]
	\begin{center}
	\includegraphics[width=0.95\columnwidth]{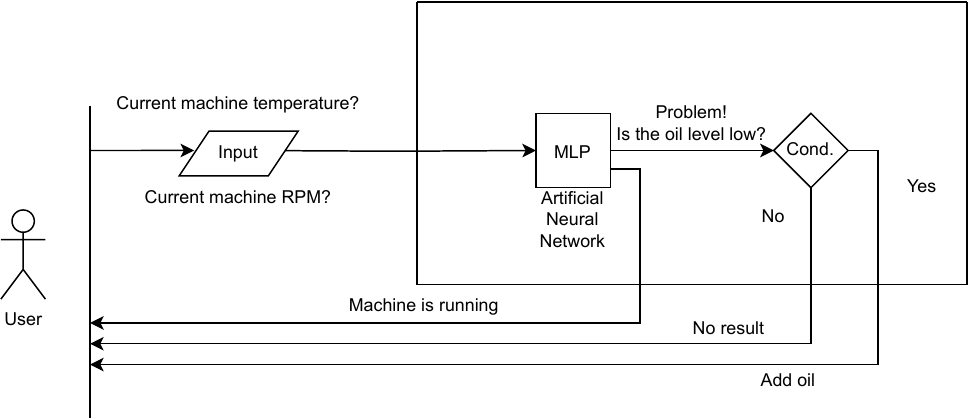}
	\end{center}
    \caption{Graphical representation of the \textit{QRind} running example.}
	\label{fig:QRind_graph}
 \vspace{-0.2cm}
\end{figure}

To better understand QRind and to explain how it works, a concrete and simple example has been used.
A high-level graphical representation of the related program is depicted in Fig.~\ref{fig:QRind_graph}. 
The very same program, described in the intermediate language, is reported in Fig.~\ref{fig:inter}.

The program initially asks the user the current \textit{temperature} and \textit{rotational speed} in revolutions per minute (RPM) of a hypothetical industrial machine. 
These input parameters are stored in register array elements
\texttt{R1[0]} and \texttt{R1[1]}, in lines \texttt{(2)} and \texttt{(4)} of Fig.~\ref{fig:inter}, respectively.
Register \texttt{R1} is then fed as input of an MLP.
After its execution (run-forward operation), if the output of the MLP is less than a given threshold, 
the message ``\textit{Machine is running}'' is printed in line \texttt{(9)}.
Such threshold is stored in register $\texttt{R0}$ and is set to $0.5$ by means of a \texttt{SET} instruction in line \texttt{(0)}.
Otherwise, if the output of the MLP is greater than or equal to $0.5$, the instruction in line \texttt{(8)} performs a conditional jump to line \texttt(11) to print a warning message (``\textit{Problem! Is the oil level fine?}'') and ask additional information to try solving the problem.

Based on the answer to the latter question, which is stored in register \texttt{R3} by the \texttt{INPUT} instruction on line \texttt{(12)}, a decision block is used to print the final message. 
In particular, the decision block that is mapped on the \texttt{TREECONDITION} instruction on line \texttt{(13)} prints ``\textit{Add oil}'' in the case the oil level is low, otherwise it prints nothing.

\begin{figure}[]
\begin{center}
\footnotesize
\begin{alltt}
(0)  SET R0 TO 0.5   # Set threshold
(1)  PRINT "Current machine temperature (°C)?"
(2)  INPUT FLOAT INTO R1[0]
(3)  PRINT "Current machine RPM?"
(4)  INPUT FLOAT INTO R1[1]
(5)  MLINPUT MLP 2 FROM R1
(6)  NNLAYER SIGMOID 1 FLOAT32 0.01, 0.001, -1.5
(7)  MLOUTPUT INTO R2
(8)  TREECONDITION R2 >= R0 (11)
(9)  PRINT "Machine is running"
(10) TREEJUMP (15)                     # exit
(11) PRINT "Problem! Is the oil level low?"
(12) INPUT BOOL INTO R3
(13) TREECONDITION R3 == FALSE (15)     # exit
(14) PRINT "Add oil"
\end{alltt}
\end{center}

  \caption{Intermediate representation of the \textit{QRind} running example.}
  \label{fig:inter}
\vspace{-0.2cm}
\end{figure}

It is worth pointing out that the intermediate language does not define an \textit{exit} instruction that stops program execution. 
This instruction is essential for QRind, and must be used whenever a final branch of the decision tree is reached. 
However, we decided to implicitly encode it into jump instructions. 
In particular, when a jump is performed to a point after the last instruction of the program, that jump is interpreted as the end of the program execution. 
In the example, this operation is performed in lines \texttt{(10)} and \texttt{(13)}.

Regarding the MLP model, it is encoded in Fig.~\ref{fig:inter} from line \texttt{(5)} to line \texttt{(7)}. 
In detail, line \texttt{(5)} maps the \texttt{R1} register to the two inputs of the ML model, which is defined as an MLP. 
The \texttt{NNLAYER} instruction defines an MLP layer of dimension $2 \times 1$ (i.e., two weights plus a bias, encoded as float32) that makes use of the \textit{sigmoid} activation function. 
Starting from weights, the output of the neural network is obtained through (\ref{eq:ann}) by means of the following computation:
\begin{equation}
out = \frac{1}{1+e^{-(\mathrm{R1[0]}\cdot 0.01 + \mathrm{R1[1]}\cdot 0.001 - 1.5  )}}
\end{equation}
where $\frac{1}{1+e^{-x}}$ is the sigmoid function. 
For instance, when the inputs are $\unit[60]{^\circ C}$ and $\unit[1000]{RPM}$, the output is $out=0.53$, which is classified as an operational problem. 
Finally, the output of the ML model is mapped to register \texttt{R2} by the \texttt{MLOUTPUT} command on line \texttt{(7)}.

Of course, in a real industrial-grade implementation of the whole toolchain, following the execution chain 
(on the right side of Fig.~\ref{fig:generationExecutionChain}), the reconstructed intermediate code of QRind
(which will appear similar to the one reported in Fig.~\ref{fig:inter}) will be executed on a virtual machine, and possibly mapped onto a graphical interface as done for QRtree.
To provide some hints on the practical capabilities of QRind, a single QR code can embed an MLP with $20$ inputs, one hidden layer with $25$ neurons, and one output, plus some additional input/output instructions.
In this case, ANN coefficients take $1102$ and $2204$ bytes, when encoded as float16 and float32, respectively.

\section{Conclusions}
\label{sec:conc}
A new eQR code dialect named QRind was defined and presented, which
supports the implementation of decision trees and also integrates machine learning algorithms.
This permits it to be applied to a variety of industrial automation contexts.

A simple example is also provided, which includes all the main instructions made available by QRind and shows how they can be employed.
This is a reasonable starting point for enabling the full implementation of the dialect. 
As future work, we will define precisely how to optimally map every instruction to the corresponding binary representation, 
i.e., occupying as little space as possible. 
This process is relatively straightforward for anybody who is knowledgeable about formal languages and translators, and can be performed using the same steps followed for the QRtree dialect.

\bibliographystyle{IEEEtran}
\bibliography{bibliography}

\end{document}